\documentclass[12pt]{article}
\usepackage{latexsym,amsmath,amsfonts,amssymb,eepic}
\usepackage{graphicx}
\newtheorem{theorem}{Theorem}

\newtheorem{lemma}{Lemma}

\newenvironment{proof}[1][Proof]{\textbf{#1.} }{\ \rule{0.5em}{0.5em}}
\begin{document}
\title{A polynomial time $\frac 3 2$ -approximation algorithm for the vertex cover
 problem on a class of graphs}
\author{Qiaoming Han
\thanks{Department of Management Science and
Engineering, Nanjing University, P.R.China. Email: qmhan@nju.edu.cn.
Research supported in part by Chinese NNSF grant, and the Return-from-Abroad foundation of MOE China.}  \and
Abraham P. Punnen\thanks{Department of Mathematics, Simon Fraser
University, 14th Floor Central City Tower, 13450 102nd Ave., Surrey,
BC V3T5X3, Canada. E-mail: apunnen@sfu.ca} \and Yinyu Ye
\thanks{Department of Management Science and Engineering,
School of Engineering, Stanford University, Stanford, CA 94305,
USA. Email:yinyu-ye@stanford.edu}}
\date{}
\maketitle
\begin{abstract}
We develop a polynomial time $\frac{3}{2}$-approximation algorithm
to solve the vertex cover problem on a class of graphs satisfying
a property called ``active edge hypothesis''.
The algorithm also guarantees an
optimal solution on specially structured graphs. Further, we give
an extended algorithm which guarantees a vertex cover $S_1$ on an
arbitrary graph such that $|S_1|\leq \frac{3}{2} |S^*|+\xi$ where
$S^*$ is an optimal vertex cover and $\xi$ is an error bound
identified by the algorithm. We obtained $\xi = 0$ for all the
test problems we have considered which include specially
constructed instances that were expected to be hard. So far we could not construct a graph that gives $\xi \not= 0$.
\end{abstract}
\par

\newpage
\section{Introduction}
Let $G=(V, E)$ be an undirected graph on the vertex set
$V=\{1,2,\ldots ,n\}$. A {\it vertex cover} of $G$ is  a subset $S$
of $V$ such that each edge of $G$ has at least one endpoint in $S$.
The {\it vertex cover problem} (VCP) is to compute a vertex cover of
smallest
cardinality in $G$. VCP is a classical NP-hard problem.

\vskip 5pt

It is well known that an optimal vertex cover of a graph can be
approximated within a factor of 2 in polynomial time by taking all
the vertices of a maximal (not necessarily maximum) matching in the
graph or rounding the LP relaxation solution of an integer
programming formulation~\cite{nt75}. There has been considerable
work (see e.g. survey paper \cite{hoch}) on the problem over the
past 30 years on finding a polynomial-time approximation algorithm
with an improved performance guarantee.  It is known that computing
a $\delta$-approximate solution in polynomial time for VCP is
NP-Hard for any $\delta\leq 10\sqrt{5}-21\simeq 1.36$~\cite{dinur},
which improved the previously known non-approximability bound of
$\frac 7 6$ in \cite{hastad97}. In fact, no polynomial-time
$(2-\epsilon)$-approximation algorithm is known for VCP for any
constant $\epsilon>0$. Under the assumption of unique game
conjecture~\cite{harb,khot1,khot} many researchers believe that a
polynomial time $2-\epsilon$ approximation algorithm is not possible
for VCP. The current best known bound on the performance ratio of a
polynomial time approximation algorithm for VCP is
$2-\Theta(\frac{1}{\sqrt{\log n}})$
 ~\cite{kara05},  which improved the
previously known ratio of $2-\frac{\log\log n}{2\log n}$
\cite{bar85,monien85}.  Halperin \cite{halperin02} showed that an
approximation ratio of $2-\frac{2\log\log \Delta}{\log \Delta}$ can
be obtained with the semidefinite programming (SDP) relaxation of
VCP where $\Delta$ is the maximum degree of $G$. Other
 SDP-relaxations of the VCP were studied in
 \cite{charikar,goemans98}. On four colorable graphs, a $\frac{3}{2}$-approximate solution can be identified
 in polynomial time.
 Recently Asgeirsson and Stein~\cite{stein,stein1} reported extensive
experimental results using a heuristic algorithm which obtained no
worse than $\frac{3}{2}$-approximate solutions for all the test
problems they considered.

\vskip 5pt
A natural integer programming formulation of VCP can be described as follows:\\
\begin{equation}
\label{vc}(VC) \hskip 15pt
\begin{array}{lcl}
&\min & \sum_{i=1}^nx_i \\
&s.t. & x_i+x_j\ge 1, (i,j)\in E,\\
 &    & x_i\in \{0, 1\}, i=1,2,\cdots,n.
\end{array}
\end{equation}
Let $\bar{x}=(\bar{x}_1,\bar{x}_2,\ldots ,\bar{x}_n)$ be an optimal
solution  to (\ref{vc}). Then  $R=\{i\ | \ \bar{x}_i=1\}$ is an
optimal vertex cover of graph $G$. The linear programming relaxation
of the above integer program, denoted by LP, is given by relaxing the integrality constraints to $x_i\ge 0, i=1,2,\cdots,n$.

\vskip 5pt

Any vertex cover must contain at least $s+1$ vertices of an odd
cycle of length  $2s+1$. This motivates  the following extended
linear programming (ELP) relaxation  of the VCP:
\begin{equation}
\label{elp}(ELP) \hskip 10pt
\begin{array}{lcl}
&\min & \sum_{i=1}^nx_i \\
&s.t. & x_i+x_j\ge 1, (i,j)\in E,\\
 &    & \sum_{i\in \omega_k}x_i\ge s_k+1, \omega_k\in \Omega,\\
  &   &  x_i\ge 0, i=1,2,\cdots,n,
\end{array}
\end{equation}
where $\Omega$ denotes the set of all odd-cycles of  $G$ and
$\omega_k\in\Omega$ contains $2s_k+1$ vertices for some integer
$s_k$. Note that even if there are an exponential number of
odd-cycles in  $G$, we know that the set of odd cycle inequalities
has a polynomial-time separation scheme and hence the ELP is
polynomially solvable. Further, it is possible to compute an optimal
basic feasible solution (BFS) of ELP in polynomial time.

\vskip 5pt

Arora, Bollob$\acute{a}$s and Lov$\acute{a}$sz \cite{arora} studied
the effect of adding odd-cycle inequalities to the LP relaxation of
the VCP. They proposed that the integrality gap of the LP with all
the odd-cycle inequalities is basically 2 in \cite{arora}.

\vskip 5pt

By solving a series of ELP relaxations on appropriately defined
graphs, we show that a $\frac 3 2$-approximation algorithm for VCP
can be obtained in polynomial time for a large class $\mathbb{F}$ of
graphs.
For all
graphs $G\in \mathbb{F}$ the integrality gap is $\frac{3}{2}$.
Further, for an arbitrary graph, we develop a polynomial time
approximation algorithm for VCP that guarantees a solution $S_1$
such that $|S_1|\leq \frac{3}{2}|S^*|+\xi$ where $S^*$ is an optimal
solution and $\xi \geq 0$ is an error bound output by the algorithm.
So far, we could not compute an explicit example of reasonable size
 where $\xi \neq 0.$

\vskip 5pt

For any graph $G$, we sometimes use the notation $V(G)$ to represent
its vertex set and $E(G)$ to represent its edge set.

\section{The Approximation Algorithm}\label{ELPA}
We first introduce some notations and
definitions. An odd cycle $\omega_1$ \emph{dominates} another odd
cycle $\omega_2$ (denoted by $\omega_1\prec \omega_2)$ if all
vertices of $\omega_1$ are contained in $\omega_2$. In this case
we also use the terminology $\omega_2$ is \emph{dominated by}
$\omega_1$. Note that an odd cycle $\omega$ is not dominated by
any other odd cycle in $G$ if and only if $\omega$ is cordless. If
$\omega_1 \prec \omega_2$ then the odd cycle constraint in ELP
corresponding to $\omega_1$ implies the odd cycle inequality
corresponding to $\omega_2$.  Two odd cycles are \emph{equivalent}
if they have the same vertex set. Note that the number of cordless
odd-cycles in graph $G$ is no more than that of triangles in the
complete graph with the same number $n$ of vertices. Thus the
number of cordless odd cycles in a graph on $n$ vertices is  $
O(n^3)$.

\vskip 5pt

Our approximation algorithm performs a series of graph reduction
operations. Let us first discuss these reductions and their inherent
properties.

\vskip 5pt

\noindent {\bf 3-cycle reduction:} This reduction was considered
earlier by many researchers including most recently by Asgeirsson
and Stein~\cite{stein,stein1}. Its properties associated with the
ELP relaxation and its power when used in conjunction with our other
reductions resulted in superior outcomes. Suppose $G$ be a graph
containing a 3-cycle. Without loss of generality assume there is a
3-cycle on vertices $\{n-2,n-1,n\}$. Let $\bar{G}=G\setminus
\{n-2,n-1,n\}$. Let $x^0=(x_1^0,x_2^0,\ldots ,x_n^0)$ be an optimal
basic feasible solution (BFS) for the ELP on $G$ with objective
function value $z(x^0)$ and $\bar{x}=(\bar{x}_1,\bar{x}_2,\ldots
\bar{x}_{n-3})$ be an optimal BFS for the ELP on $\bar{G}$ with
optimal objective function value $\bar{z}(\bar{x})$.
\begin{lemma}\label{3c1}$\bar{z}(\bar{x})\leq z(x^0)-2$.\end{lemma}
\begin{proof}Note that $x^1=(x^1_1,x^1_2,\ldots ,x^1_{n-3})$  defined by $x^1_j=x^0_j$ for
 $1\leq j \leq n-3$ is a feasible solution
to ELP on $\bar{G}$. Thus its objective function value $z(x^1)$
satisfies $\bar{z}(\bar{x}) \leq z(x^1)$. But $z(x^1)+2 \leq z(x^0)$
since $x^0_{n-2}+x^0_{n-1}+x^0_{n}\geq 2$ and the result follows.
\end{proof}

\vskip 10pt

\noindent {\bf Active edge reduction:} This reduction technique is
very powerful with some interesting properties. Let $x^0$ be an
optimal BFS for the ELP on $G$ and let $(i,j)$ be an \emph{active
edge} in $G$ with respect to the solution $x^0$, i.e.,
$x^0_i+x^0_j=1$.  Let $D_i=\{s\in V(G)\ |\ (i,s)\in E(G), s\not=j\},
D_j=\{t\in V(G)\ |\ (t,j)\in E(G), t\not=i\}$. Construct the new
graph $G^{(i,j)}$ from $G$ as follows. In graph $G,$ connect each
vertex $s\in D_i$ to each vertex $t\in D_j$ whenever such an edge is
not already present and delete vertices $i$ and $j$ with all the
incident edges. The operation of constructing $G^{(i,j)}$ from $G$
is called \emph{active edge reduction}.

\begin{lemma}\label{ll1} If an active edge $(i,j)$ is contained in
an odd cycle, say $\omega=(i,v_1,v_2,\ldots ,v_k,j)$ in $G$ then
$\sum_{j\in \omega_0}x_j^0\geq \lceil \frac{k}{2}\rceil$ where
$\omega_0$ is the vertex set $\{v_1,v_2,\ldots ,v_k\}.$\end{lemma}

The proof of this lemma is straightforward. The lemma shows that if
an odd cycle has an active edge, there is an implicit sub-odd-cycle
for the odd cycle where the solution $x^0$ satisfies this smaller
implicit odd cycle constraint. Let $z^{(i,j)}$ be the optimal
objective function value of ELP on $G^{(i,j)}$. The following lemma
provides a somewhat surprising property of the active edge
reduction.

\begin{lemma}\label{a1} If $G$ does not contain 3-cycles using arc $(i,j)$, $z^{(i,j)}\leq z(x^0)-1$.\end{lemma}
\begin{proof}Since $G$ does not contain 3-cycles using arc $(i,j)$, we have $D_i\cap
D_j=\emptyset$. We now show that
$\hat{x}=x^0\setminus\{x^0_i,x^0_j\}$ is a feasible solution to ELP
on $G^{(i,j)}.$ Note that $(i,s)$ for all $s\in D_i$ and $(j,t)$ for
all $t\in D_j$ are edges of $G$. Thus
\begin{eqnarray} x^0_i+x^0_s &
\geq 1 \mbox{ for all } s\in D_i\ ,\label{e1}\\
x^0_j+x^0_t & \geq 1 \mbox{ for all } t\in
D_j\ .\label{e2}\end{eqnarray}
Since $(i,j)$ is an active edge
$x^0_i+x^0_j=1$. Adding (\ref{e1}) and (\ref{e2}) we get
$\hat{x}_s+\hat{x}_t=x^0_s+x^0_t \geq 1$ for all $s\in D_i$ and
$t\in D_j$. Thus $\hat{x}$ satisfies all edge inequalities in the
ELP on $G^{(i,j)}.$ The addition of the new edges $(s,t)$ for
$s\in D_i$ and $t\in D_j$ probably created new odd cycles in
$G^{(i,j)}$. Any such odd cycle must be one of the following four
types:
\begin{itemize}
\item[Type 1:] Uses exactly one new edge $(s_1,t_1)$ where $s_1 \in D_i, t_1\in D_j$;
\item[Type 2:] Uses exactly two new edges of the form $(s_1,t_1), (s_2,t_1)$
where $s_1,s_2 \in D_i$ and $t_1 \in D_j$;
\item[Type 3:] Uses exactly two new edges of the form $(t_1,s_1), (t_2,s_1)$
where $s_1 \in D_i$ and $t_1,t_2 \in D_j$;
\item[Type 4:] Must contain a sub odd cycle which is of type 1,2, or
3 above.
\end{itemize}

Let $\omega_1$ be a Type 1 odd cycle in $G^{(i,j)}$.  Then
$\omega_2=\omega_1\cup\{(i,j),(i,s_1),(j,t_1)\}\setminus\{(s_1,t_1)\}$
must be an odd cycle in $G$. Then by Lemma~\ref{ll1} $\hat{x}$ must
satisfy the odd cycle inequality corresponding to $\omega_1$. Let
$\omega_3$ be a Type 2 odd cycle in $G^{(i,j)}.$ Then it is of the
form $\omega_3=\{s_1,t_1,s_2,P(s_1,s_2)\}$ where
$P(s_1,s_2)$ is a path in both $G^{(i,j)}$ and $G.$  Then
$\omega_4=\{s_1,i,s_2, P(s_1,s_2)\}$ must be an odd cycle in $G$. From
(\ref{e1}), (\ref{e2}) and  $x^0_i+x^0_j=1$, we have
\begin{eqnarray}\label{e3}
x^0_i\leq  x_t \mbox{ for all } t\in D_j\ , \\
\label{e4} x^0_j\leq x_s \mbox{ for all } s\in
D_i\ .
\end{eqnarray}
Thus, since $x^0$ satisfy the odd cycle inequality
corresponding to $\omega_4$, in view of (\ref{e3})
$\hat{x}$ must satisfy the odd cycle inequality corresponding to
$\omega_3$. The case of Type 3 odd cycles is similar to Type 2 odd
cycles and it can be verified that $\hat{x}$ satisfies these odd
cycle inequalities as well. Since $\hat{x}$ satisfies all edge
inequalities in $G^{(i,j)}$ and also satisfies all odd cycle
inequalities corresponding to odd cycles of the form Type 1, Type 2,
Type 3, and those does not use any new edge of $G^{(i,j)}$, by
dominance property, it must satisfy all Type 4 odd cycle
inequalities. Thus  $\hat{x}$ is a feasible solution to the ELP on
$G^{(i,j)}.$ Let $ \hat{z}$ be the objective function of $\hat{x}$.
Then $z^{(i,j)}\leq \hat{z}$. But $\hat{z}=z(x^0)-1$ and the result
follows.\end{proof}

\vskip 0.7cm

\begin{lemma}If $R$ is a vertex cover of
$G^{(i,j)}$ then \begin{equation*}R^* =\begin{cases} R\cup \{j\}, & \text{ if } D_i\subseteq R;\\
R\cup \{i\}, & \text{ otherwise},\end{cases}
\end{equation*}is a vertex cover of $G$.
\end{lemma}
\begin{proof} If
$D_i\subseteq R$ then all arcs in $G$ incident on $i,$ except
possibly $(i,j),$ is covered by $R$. Then $R^*=R\cup \{j\}$ covers
all arcs incident on $j$, including $(i,j)$ and hence $R^*$ is a
vertex cover in $G$. If at least one vertex of $D_i$ is not in $R$,
then all vertices in $D_j$ must be in $R$ by construction of
$G^{(i,j)}$. Thus $R\cup\{i\}$ must be a vertex cover of $G$.
\end{proof}
\vskip 9pt

\noindent \textbf{Over-active edge reduction:} An edge $(i, j)$ is over active with respect to an ELP optimal BFS $x^0$ if
$x_i^0+x_j^0\ge\frac{4}{3}$. Let
$\bar{G}^{\{i,j\}}=G\setminus \{i,j\}$, and $\bar{x}$ be an
optimal BFS for the ELP on $\bar{G}^{\{i,j\}}$ with objective function value
$\bar{z}(\bar{x})$.

\begin{lemma} \label{z0} $\bar{z}(\bar{x})\leq z(x^0)-\frac 4 3.$\end{lemma}

\noindent \textbf{$\{0,1\}$-reduction:} Let $I_0 = \{i
~:~ x^0_i=0\}$ and $I_1 = \{i ~:~ x_i^0=1\}.$ Consider the graph
$\acute{G}=G\setminus \{I_0\cup I_1\}.$ Let $\acute{x}$ be an
optimal BFS for the ELP on $\acute{G}$ with objective function value
$\acute{z}(\acute{x})$.
\begin{lemma}\label{z1} If $R$ is a vertex cover of $\acute{G}$ then
$R\cup I_1$ is a vertex cover of $G$.
Further, $\acute{z}(\acute{x})\leq z(x^0)-|I_1|.$\end{lemma}

We skip the proof of Lemma \ref{z0} and \ref{z1}, which is easy to obtain. The active
edge hypothesis discussed below is the assumption we make in the
algorithm. The algorithm guarantees a $\frac{3}{2}$-approximate
solution when this hypothesis is valid.\\

\noindent \textbf{Active Edge Hypothesis:} \textit{Let $G$ be a
graph and $x^0 = (x^0_1,x^0_2,\ldots ,x^0_n)$ be an optimal BFS of
the ELP relaxation on $G$. Then at least one of the following is
true:
\begin{enumerate}
\item $G$ contains a 3-cycle;
\item There
exists at least one active edge in $G$ with respect to the solution
$x^0$;
\item There
exists at least one over active edge in $G$ with respect to the solution
$x^0$;
\item There is at least one $x_i^0=1,~1\leq i \leq
n$.\end{enumerate}}

Let us now discuss our approximation algorithm. The algorithm
guarantees a $\frac{3}{2}$-approximate solutions when the
intermediate graphs used in the algorithm satisfies the active edge
hypothesis. The basic idea of the algorithm is very simple. We apply
3-cycle, active edge, over active edge and $\{0,1\}$ reductions repeatedly until  the
underlying ELP solution is integer, in which case the algorithm goes
to a back tracking step. Active edge hypothesis guarantees this
termination criterion for all graphs for which it is valid. If we
encounter a graph that violates the active edge hypothesis, the
algorithm is terminated. We record the vertices in the active edge
reductions step but do not determine which one to be included in the
vertex cover. In the back track step we choose exactly one of these
two vertices to form part of the vertex cover we construct. Active
edge reduction may create  new odd cycles in the graph under
consideration which in turn could result in additional 3-cycles at
later stages of the reduction steps and then 3-cycle and $\{0,1\}$
reduction steps are applied again and the whole process is continued
until we reach the back tracking step. In this step, the algorithm
computes a vertex cover for $G$ using the integer solution obtained
in the last reduction step together with all vertices removed in
3-cycle and over active edge reductions, vertices with value 1 removed in the $\{0,1\}$
reduction steps, and the selected vertices in the backtrack step
from the active edges recorded during the active edge reduction
steps. A formal description of the ELP-Algorithm is given below.

\vskip 10pt
\begin{itemize}
\item[~] {\large\bf The ELP-Algorithm}
\item[Step 1:~] \{* \textsf{Initialize} *\}\
$G_1=G, k=1$.
\item[Step 2:~] Solve the ELP relaxation of VCP on graph $G_k$. Let
$x^k=\{x_i^k : i\in V(G_k)\}$ be the resulting optimal BFS with
optimal objective function value $f^k$.
\item[Step 3:~] \{* \textsf{Reduction operations}
*\}\ $\Delta_k=\emptyset$, $I_{k,1}=\emptyset,$ $(i_k,j_k)=\emptyset$, $(\bar{i}_k, \bar{j}_k)=\emptyset$.
\begin{enumerate}
\item \{* \textsf{\{0,1\}-reduction} *\} Let $I_{k,0}=\{i\ | \ x_i^k=0\}, \
I_{k,1}=\{ i\ | \ x_i^k=1\}, $ and $I_k=I_{k,0}\cup I_{k,1}$.
\textbf{If} $V(G_k)\setminus I_{k}=\emptyset$ \textbf{goto}
step 4 \textbf{else}  $G_k=G_k\setminus I_k$ \textbf{endif}
\item \{* \textsf{3-cycle reduction} *\} \textbf{If} $G_k$ has 3-cycles \textbf{then} \\
     \mbox{ }Choose a 3-cycle $\Delta_k$. Set $G_{k+1}=G_k\backslash \Delta_k$;
     $k=k+1,$ \textbf{goto} Step 2 \textbf{endif}
\item \textbf{If} $G_k$ has neither active edges  nor over-active edges \textbf{then} k=k+1 and \textbf{goto} Step 2 \textbf{endif}
\item \{* \textsf{active edge reduction} *\} \ \textbf{If} $G_k$ has active edges \textbf{then}
 Choose an active
edge $(i,j)$. Let $G_{k+1}=G_k^{(i,j)}$ where $G_k^{(i,j)}$ is the
graph obtained from $G_k$ using  active edge reduction operation.
Let $i_k=i, j_k=j$; $k=k+1$
\textbf{goto} Step 2 \textbf{endif}
\item \{* \textsf{over-active edge reduction} *\}\ \textbf{If} $G_k$ has over-active edges \textbf{then} \\
     \mbox{ } Choose an over-active edge $(i, j)$. Set $G_{k+1}=G_k\backslash \{i,j\}$, and $\bar{i}_k=i, \bar{j}_k=j$;
     $k=k+1,$ \textbf{goto} Step 2 \textbf{endif}
\end{enumerate}
\item[Step 4:~] L=k.  Let $S_L=I_{L,1}$. \textbf{If} $k=1$ \textbf{then} output
$S_1$ and STOP \textbf{endif}
\item[Step 5:~] \{* \textsf{Backtracking to construct a solution} *\}\\
Let $S_{k-1}=S_{k}\cup I_{k-1,1}$,\\
\textbf{If} $\triangle_{k-1}\not=\emptyset,$ \textbf{then}
$\mbox{ ~ }S_{k-1}=S_{k-1}\cup \triangle_{k-1}$ \textbf{endif}\\
\textbf{If} $(i_{k-1},j_{k-1})\neq \emptyset$ \textbf{then}
$\mbox{ ~ }S_{k-1}=S_{k-1}\cup R^*$, where
\begin{equation*}R^* =
\begin{cases} j_{k-1}, & \text{ if } D_{i_{k-1}}\subseteq S_k;\\
i_{k-1}, & \text{ otherwise},\end{cases}
\end{equation*}
$\mbox{ ~ }$ and $D_{i_{k-1}}=\{s  : (i_{k-1},s)\in G_{k-1},i_{k-1}\ne
j_{k-1}\}$ \textbf{endif}\\
\textbf{If} $(\bar{i}_{k-1}, \bar{j}_{k-1})\not=\emptyset,$ \textbf{then}
$\mbox{ ~ }S_{k-1}=S_{k-1}\cup \{\bar{i}_{k-1}, \bar{j}_{k-1}\}$ \textbf{endif}\\
$ k = k-1$,\\ \textbf{If} $k\neq 1$ \textbf{then} \textbf{goto}
beginning of step 5 \textbf{else} output $S_1$ and STOP
\textbf{endif}
 \end{itemize}

\vskip 5pt
\begin{theorem}
\label{conclusion} Under the active edge hypothesis on $G_k$ for $k=1, \cdots, L-1$, the ELP
Algorithm correctly identifies a $\frac 3 2$-approximate solution
$S_1$ for the vertex cover problem on $G$ in polynomial time.
\end{theorem}
\begin{proof} Note that if $I_{k,1}=\emptyset$ at any iteration $k$, then by the active edge hypothesis, $G_k$
must contain an active edge or an over-active edge, or it must contain a 3-cycle. Thus in
each execution of Step 2, at least one node is removed. Thus the
algorithm executes Step 2 $O(n)$ times and the backtrack step takes
at most $n$ iterations where $n=|V(G)|$. The complexity of Step 2 is
polynomial since the LP can be solved in polynomial time. Thus it
can be verified that
the complexity of the algorithm is polynomial.

\vskip 5pt

To establish the validity of the algorithm, note that $S_k$ is a
vertex cover for graph $G_k$ for $k=L,\cdots,1$. In particular,
$S_1$ is a vertex cover for the graph $G$. Let $f^k$ be the
objective function value at the LP solution identified in Step 2 at
the kth execution of the step. Then $f^L=d^L=|I_{L,1}|=|S_L|.$
Further, from Lemma~\ref{3c1}, \ref{a1}, \ref{z0} and \ref{z1},
\begin{equation}
\label{objective}
f^{k+1}\le f^k-d_k, \ \ k=1, 2,\cdots, L-1,\\
\end{equation}
where
\[
d_k=\left\{
\begin{array}{ll}
|I_{k,1}|+2, & \mbox {if }\triangle_k\not=\emptyset;\\
|I_{k,1}|+1, & \mbox{if }(i_{k},j_k)\neq \emptyset;\\
|I_{k,1}|+\frac{4}{3}, & \mbox{if }(\bar{i}_{k},\bar{j}_k)\neq \emptyset;\\
|I_{k,1}|, &
\mbox{otherwise}.
\end{array}  \right.
\]
Adding inequality (\ref{objective}) for  $k=1$ to $L$, we get that
\( \Sigma_{k=1}^Ld_k \le f^1. \) Note that $|S_{k}|-|S_{k+1}|$ is
the number of vertices added to the  vertex cover constructed for
$G_{k+1}$ to obtain the vertex cover constructed for $G_{k}$ in the
$k$'th iteration of the backtrack step. Note that $|S_k|-|S_{k+1}|
\leq |I_{k,1}|+3$ if 3-cycle reduction is used to construct
$G_{k+1}$ from $G_{k}$, $|S_k|-|S_{k+1}| \leq |I_{k,1}|+1$ if active
 edge reduction is used to construct $G_{k+1}$ from $G_{k}$, $|S_k|-|S_{k+1}| \leq |I_{k,1}|+2$ if over-active
 edge reduction is used to construct $G_{k+1}$ from $G_{k}$, and $|S_k|-|S_{k+1}| \leq |I_{k,1}|$
 if only $\{0,1\}$-reduction is used to construct $G_{k+1}$ from $G_k$. Thus we
have $|S_{k}|-|S_{k+1}|\le \frac 3 2 d_k$ for $k=L-1,\cdots,1.$ Now,
\[
|S_1|=|S_L|+\Sigma_{k=1}^{L-1}(|S_{k}|-|S_{k+1}|)\le\frac 3 2
\Sigma_{k=1}^Ld_k \le \frac 3 2 f^1\le \frac 3 2|S^*|,
\]
where $S^*$ is an optimal vertex cover of $G$.\end{proof}

\vskip 10pt

Let us now consider a class of graphs where the active
edge hypothesis is true. Let $C$ be a cycle in $G$. The incidence vector of $C$ is the
$n$-vector $\tau_c= (\tau_c(1),\tau_c(2),\ldots ,$ $\tau_c(n) )$
where
\begin{equation}\tau_c(i) =\begin{cases} 1, & \text{ if $i\in V(C)$};\\
0, & \text{ otherwise}.\end{cases}\end{equation}
\noindent Note that equivalent cycles have the same incidence
vector. A collection $C=\{C_1, C_2, \ldots, C_p\}$ of odd cycles
in $G$ is said to be linearly independent if their incidence
vectors are linearly independent.


\begin{theorem}Let $G$ be a graph containing triangles or has less than $|V(G)|$ independent
cordless odd cycles, then $G$ satisfies the active edge
hypothesis.\end{theorem}

\begin{center}
\begin{figure}
  \includegraphics[width=7.0cm]{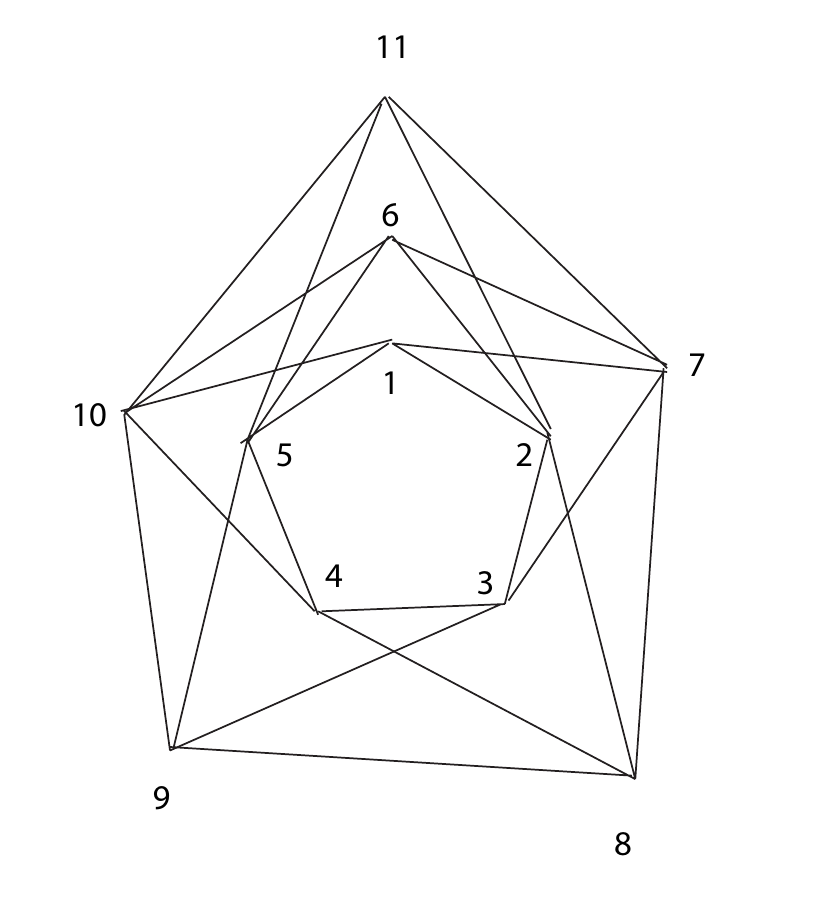}  \ \  \includegraphics[width=7.0cm]{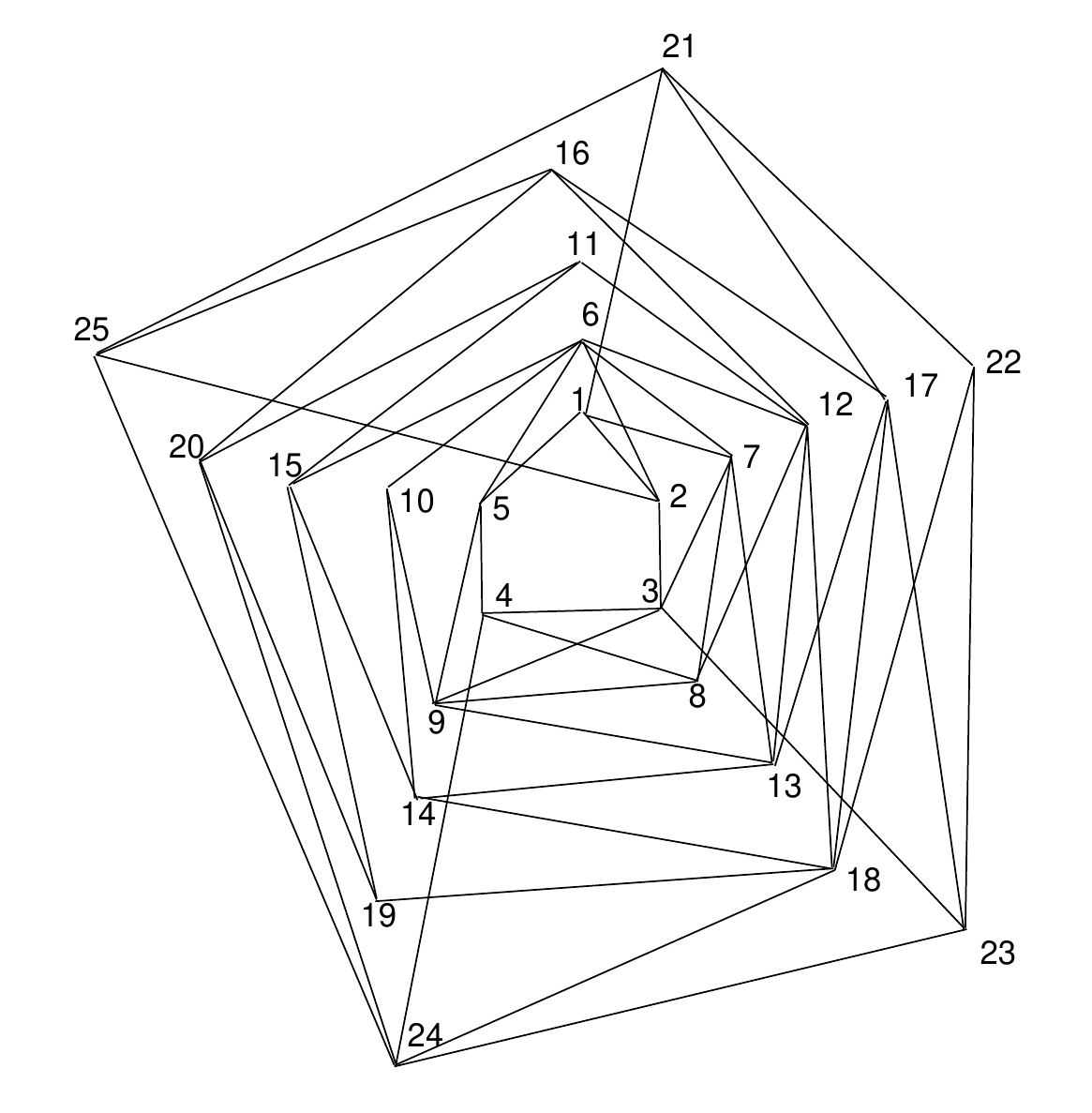}
  \caption{Left graph on 11 nodes has only 7 linearly independent odd holes.
  Right graph on 25 nodes has 25 linearly independent odd holes. Both graphs have no 3-cycles.}\label{fig1}
\end{figure}
\end{center}

Left graph in Figure 1 below gives an example of $\bar{G}$ on 11
nodes with more than 11 cordless odd cycles but only 7 of them are
independent. Active edge hypothesis is true on this graph or any
subgraph of it or a super graph of it obtained by adding 3-cycles.
However, it is possible to construct graphs with $n$ nodes and $n$
independent odd cycles and
have no 3-cycles.
Right graph in Figure 1 below gives such a graph on 25 nodes with
no 3-cycles and 25 independent 5-cycles. The vector with $x_i=\frac{3}{5}$
for $i=1,2,\ldots 25$ is an optimal BFS of the ELP on this graph.
If we encounter this BFS on this graph in the ELP reduction
algorithm, we terminate with the flag that ``active edge
hypothesis failed''. It may be noted that there are alternative
optimal BFS to this ELP relaxation which is integer. In fact
solving this  ELP using LINDO generated integer optimal solution
\{$x_i=0$, if $i=2,5,8,10,13,15,16,18,21,23;\ x_i=1$, otherwise\}
and not the fractional optimal solution we constructed above. Thus
even if we encounter a situation where the active edge hypothesis
is not satisfied in the algorithm, one may look for an active edge
in an alternative optimal solution. Such a solution can be
explored by forcing one of edge inequalities to be equality in the
ELP and solving this modified ELP for each edge. At any stage, if
the objective function value is not increased, then we have an
alternative ELP solution with an active
edge and the active edge reduction can be carried out.\\

\section{Potpourri Extensions}

Let us now discuss various techniques to handle the situation where
the active edge hypothesis fails. These techniques provides minor
improvements on the performance of the algorithm.

\vskip 5pt

\noindent \textbf{Random edge reduction:}  Remove an edge
$(i,j)$ from $G$ along with its two incident nodes.

\vskip 5pt

Without loss of generality assume $(i,j)=(n-1,n)$ and let
$\bar{G}=G\setminus \{n-1,n\}.$  Let $x^0=(x_1^0,x_2^0,\ldots
,x_n^0)$ be an  optimal BFS for the ELP on $G$ with objective
function value $z(x^0)$ and $\bar{x}=(\bar{x}_1,\bar{x}_2,\ldots
\bar{x}_{n-2})$ be an optimal BFS for the ELP on $\bar{G}$ with
optimal objective function value $\bar{z}(\bar{x})$.

\begin{lemma}\label{3c5}$\bar{z}(\bar{x})< z(x^0)-1$.\end{lemma}

Reduction operations in Algorithm ELP can easily be modified to
incorporate the random edge reduction step. Unlike the active edge
reduction, which chooses exactly one node from an active edge, the
random edge reduction takes both nodes of the edge selected
randomly. But the optimal vertex cover does not necessarily contain
both these nodes and may contain only one of them. This is a
2-approximation. The active edge reduction and $\{0,1\}$-reduction
however preserve optimality. Thus for each node selected in a
$\{0,1\}$-reduction step or an active edge reduction step, we can
perform one random edge reduction in the algorithm and still
preserve the $\frac{3}{2}$-approximation guarantee. To improve the
probability of a $\frac{3}{2}$-approximation guarantee, we want to
make sure the total number of nodes collected in active edge
reduction step and $\{0,1\}$-reduction step to be as large as
possible. So it is better to perform an active edge reduction step
in the ELP reduction algorithm before the three cycle reduction. To
achieve this we want to make sure $(i,j)$ is not part of a 3-cycle
in $G_k$, otherwise Lemma \ref{a1} is not valid. Fortunately, this
is true since if $(i,j)$ is active and is part of a 3-cycle, the
third node will have a value 1 in the ELP optimal solution and the
$\{0,1\}$ reduction step would have removed this node. Consider the
enhanced ELP Algorithm where
Step 3 is replaced by:\\

\begin{itemize}
\item[Step 3:~] \{* \textsf{Reduction operations}
*\}\ $\Delta_k=\emptyset$, $I_{k,1}=\emptyset,$ $(i_k,j_k)=\emptyset$, $(\bar{i}_k, \bar{j}_k)=\emptyset$, $(\hat{i}_k, \hat{j}_k)=\emptyset$.
\begin{enumerate}
\item Let $I_{k,0}=\{i\ | \ x_i^k=0\}, \ I_{k,1}=\{ i\ | \
x_i^k=1\}$,\\\textbf{If} $|I_{k,1}|\neq \emptyset$ or $G_k$ has an
active edge, \textbf{goto} Step 3 (3) \textbf{endif}
 \item \{*
\textsf{Exploring alternate optimal BFS for active edge} *\}
$E= E(G_k)$, T=0,\\
\textbf{while} $E\neq \emptyset$ \textbf{do}
Choose an edge $(i,j)\in E$.
Solve the ELP on $G_k$ with the edge inequality corresponding to
$(i,j)$ replaced by an equality. Let $\tilde{x}$ be the optimal
BFS obtained with the objective function value $\tilde{f}$.\\
\textbf{If} $\tilde{f}=f^k$ \textbf{then} $x^k=\tilde{x}, T=1$ and \textbf{goto} Step 3(3)
\textbf{else} $E=E\setminus \{(i,j)\}$ \textbf{endif} \\ \textbf{endwhile}\\
\textbf{If} T=0, \textbf{goto} Step 3(5) \textbf{endif}\item \{*
\textsf{\{0,1\}-reduction} *\}  $I_k=I_{k,0}\cup I_{k,1}$.\\
\textbf{If} $V(G_k)\setminus I_{k}=\emptyset$, \textbf{goto}
step 4, \textbf{else}  $G_k=G_k\setminus I_k$ \textbf{endif}
\item \{* \textsf{active edge reduction} *\}  \ \textbf{If} $G_k$ has active edges \textbf{then} Choose an active
edge $(i,j)$. Let $G_{k+1}=G_k^{(i,j)}$ where $G_k^{(i,j)}$ is the
graph obtained from $G_k$ using  active edge reduction operation.
Let $i_k=i, j_k=j$; $k=k+1$ \textbf{goto} Step 2 \textbf{endif}
\item \{* \textsf{3-cycle reduction} *\} \textbf{If} $G_k$ has 3-cycles \textbf{then} \\
     \mbox{ }Choose a 3-cycle $\Delta_k$. Set $G_{k+1}=G_k\backslash \Delta_k$;
     $k=k+1,$ \textbf{goto} Step 2 \textbf{endif}
\item \{* \textsf{over-active edge reduction} *\}\ \textbf{If} $G_k$ has over-active edges \textbf{then} \\
     \mbox{ } Choose an over-active edge $(i, j)$. Set $G_{k+1}=G_k\backslash \{i,j\}$, and $\bar{i}_k=i, \bar{j}_k=j$;
     \mbox{ } $k=k+1,$ \textbf{goto} Step 2 \textbf{endif}
\item \{* \textsf{random edge reduction} *\}\ \textbf{If} the active
edge hypothesis does not hold for $G_k$ \textbf{then} choose
any edge $(i,j)$.
Let $G_{k+1}=G_k\backslash \{i,j\}$, and $\hat{i}_k=i, \hat{j}_k=j$; $k=k+1$ \textbf{goto} Step 2 \textbf{endif}
\end{enumerate}
\end{itemize}

Let $I_1=\cup_{k=1}^L I_{k,1}$ and $\eta, \gamma, \delta, \sigma$ be the
number of active-edge reductions, random-edge reductions, 3-cycle reductions and over-active edge reductions,
 respectively,  performed in the enhanced ELP
 algorithm. Let $\beta=|I_1|+\eta$, $\alpha = \max\{0,
\gamma - \beta\}$ and $\lambda = \gamma+\delta+\frac{2}{3}\sigma$.

\begin{lemma} The enhanced ELP computes a vertex cover $S_1$ on $G$ in
polynomial time such that $|S_1|\leq \frac{3}{2}|S^*|+\frac{\alpha}{2}$, where
$S^*$ is an optimal vertex cover on $G$. Further, $|S_1|\leq
|S^*|+\lambda.$\end{lemma}

Thus when $\alpha = 0$ the enhanced ELP algorithm computes a
$\frac{3}{2}$-approximate solution. When $\lambda = 0$ the
enhanced ELP algorithm computes an optimal solution. Recall that
$f^1$ is the optimal objective function value of ELP on $G$. Note
that $S_1$ has at most $\lambda$ extra nodes compared to an
optimal vertex cover. Thus if $\lambda \leq \frac{f^1}{2}$ then,
$$ |S_1|\leq |S^*|+\lambda \leq |S^*|+ \frac{f^1}{2} \leq
|S^*|+\frac{|S^*|}{2}=\frac{3}{2}|S^*|.
$$
 Let $\xi =
 \min\{\frac{\alpha}{2}, \max\{0, \lambda-\frac{f^1}{2}\}\}$.

 \begin{theorem} The enhanced ELP algorithm computes a vertex cover $S_1$ on $G$ in
polynomial time such that $|S_1|\leq \frac{3}{2}|S^*|+\xi$, where $S^*$
is an optimal vertex cover on $G$.\end{theorem}

\section{Conclusion} In this paper, we presented a polynomial time
approximation algorithm that computes a vertex cover $S_1$ such that
$|S_1|\leq \frac{3}{2}|S^*|+\xi$, where $S^*$ is an optimal vertex
cover and $\xi$ is an error factor identified by the algorithm. In
all the examples we constructed $\xi$ turned out to be zero.
It would be interesting
to compute explicit examples where $\xi\ne 0$.

\vskip 5pt

It seems that the ELP Algorithm may not guarantee a $\frac 3
2$-approximate solution for VCP on all graphs, since it is based on
the optimal objective function value of the ELP relaxation and the
integrality gap of the ELP is basically 2 \cite{arora}. The proof
in~\cite{arora} is probabilistic in nature and establishes existence
of a graph for which integrality gap is 2. No constructive proof of
this is known. The operation of active edge reduction is crucial to
our algorithm. Let $x^0$ be an optimal solution to the ELP problem,
an edge $(i, j)$ is said to be a {\it small edge} with respect to
$x^0$ if $x_i^0+x_j^0=\min\{x_r^0+x_s^0\ |\ (r, s) \in E(G)\}$. If
an active edge exists in $G$ with respect to $x^0$, then it will be
a small edge.
If $G$ has no 3-cycles and $(i,j)$ is a small edge, one may be tempted to
conjecture that
 there exists an optimal vertex cover $V^0$ of $G$ containing
only one of the nodes in $\{i,j\}$. It turns out that this is true
for a large class of graphs. If it is true in general, then it leads
to a polynomial time $\frac 3 2$-approximation algorithm  for VCP on
any graph $G$.

\begin{figure}\begin{center}
  \includegraphics[width=9.0cm]{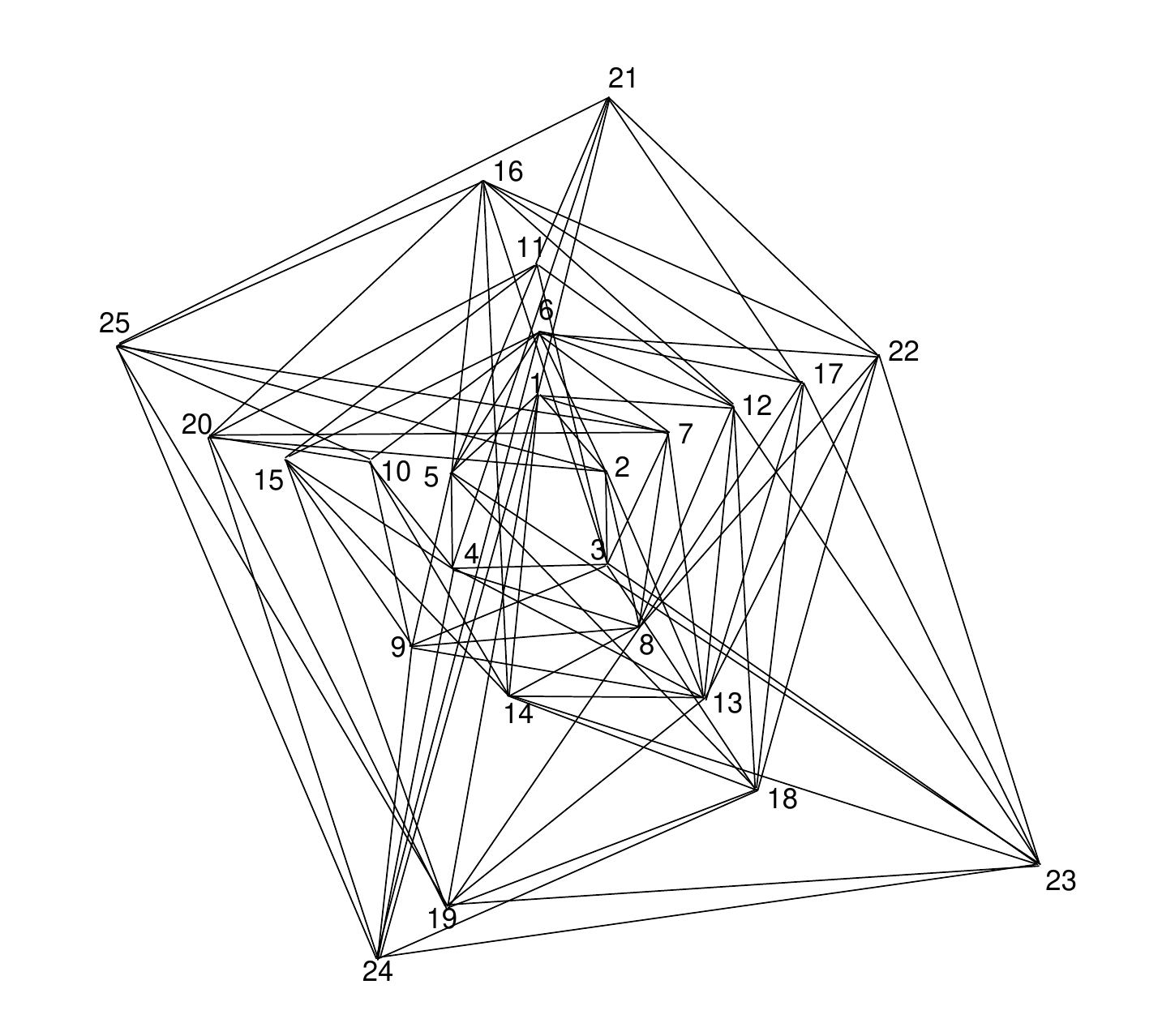}
  \caption{A graph  on 25 nodes with no 3-cycles such that both end nodes of all small edges
   are in all optimal vertex covers.}
  \label{fig2}
\end{center}\end{figure}
However, we have a very interesting counter example (See Figure
\ref{fig2}) for this establishing that such a claim is not
necessarily true. There are  five different optimal vertex covers
for the graph given in Figure \ref{fig2} which are listed below:
\[{\footnotesize
\begin{array}{ccccccccccccccccccccccccc}
  1 & 2& 3& 4& 5& 6& 7& 8& 9& 10 &11& 12& 13& 14 &15& 16& 17& 18& 19& 20& 21& 22& 23& 24& 25 \\ \hline
     1 & 0 & 1 & 0 & 1 & 1 & 0 & 1 & 0 & 1 & 0 & 1 & 1 & 0 & 1 & 1 & 0 & 1 & 0 & 1 & 1 & 0 & 1 & 1 & 1\\
     1 & 0 & 1 & 0 & 1 & 1 & 0 & 1 & 0 & 1 & 1 & 0 & 1 & 0 & 1 & 1 & 0 & 1 & 0 & 1 & 1 & 0 & 1 & 1 & 1\\
     1 & 0 & 1 & 1 & 0 & 1 & 0 & 1 & 1 & 0 & 1 & 0 & 1 & 1 & 0 & 1 & 0 & 1 & 1 & 1 & 1 & 0 & 1 & 0 & 1\\
     1 & 0 & 1 & 1 & 0 & 1 & 0 & 1 & 1 & 0 & 1 & 0 & 1 & 1 & 1 & 1 & 0 & 1 & 0 & 1 & 1 & 0 & 1 & 0 & 1\\
     1 & 0 & 1 & 1 & 0 & 1 & 0 & 1 & 1 & 1 & 1 & 0 & 1 & 0 & 1 & 1 & 0 & 1 & 0 & 1 & 1 & 0 & 1 & 0 & 1
\end{array}}
\]

The unique optimal solution $x^0=(x_1^0, x_2^0,\cdots,x_n^0)$ to ELP is given by
$x^0_i=\frac{3}{5}$ for all~$i$. Thus any edge is a small edge. If
the small edge is selected as any of the following: $(1, 21), (3,
16),  (3, 18), (3, 23)$, $(16, 20), (16, 25),  (21, 25)$, both of
their incident nodes are in all optimal vertex covers.

\vskip 5pt

Note that this graph is maximal without 3-cycles on 25
nodes, in the sense of that any additional edge will result in a
3-cycle in the graph. The graph discussed above does not satisfy
active edge hypothesis. Nevertheless,   the ELP Algorithm with
extensions discussed in Section 3 guarantees a $\frac 3
2$-approximate
solution for this graph, since only two random edge reductions are
 needed by following appropriate general rules that selects
 $(24, 25), (22, 23)$ for the operation of random edge reduction. \\

\indent{\large\bf Acknowledgement:} This work was partially
supported by an NSERC discovery grant awarded to Abraham P. Punnen.
The first author would be very grateful to Dr. Jiawei Zhang and Dr.
Donglei Du for several helpful discussions.

{ \small 


}
\end{document}